\documentclass{sig-alternate-per-modified}
\usepackage{booktabs}
\usepackage[hidelinks]{hyperref}
\providecommand{\Description}[1]{}

\begin{document}
\sloppy

\title{Alert or Noise? Reducing False Positives with Active Behavioral Analysis for Cloud Security}

\author{
Dikshant \\
7-Eleven, Bengaluru, Karnataka, India \\
\texttt{27dikshant@gmail.com}
\and
Geetika Verma \\
RMIT University Melbourne, Australia \\
\texttt{geetika.verma@rmit.edu.au}
}
\maketitle
\begin{abstract}
Rule-based cloud security posture management (CSPM) solutions are known to produce a lot of false positives based on the limited contextual understanding and dependence on static heuristics testing. This paper introduces a validation-driven methodology that integrates active behavioral testing in cloud security posture management solution(s) to evaluate the exploitability of policy violations in real time. The proposed system employs lightweight and automated probes, built from open-source tools, validation scripts, and penetration testing test cases, to simulate adversarial attacks on misconfigured or vulnerable cloud assets without any impact to the cloud services or environment. For instance, cloud services may be flagged as publicly exposed and vulnerable despite being protected by access control layers, or secure policies, resulting in non-actionable alerts that consumes analysts time during manual validation. Through controlled experimentation in a reproducible AWS setup, we evaluated the reduction in false positive rates across various misconfiguration and vulnerable alerts. Our findings indicate an average reduction of 93\% in false positives. Furthermore, the framework demonstrates low latency performance. These results demonstrate a scalable method to improve detection accuracy and analyst productivity in large cloud environments. While our evaluation focuses on AWS, the architecture is modular and extensible to multi-cloud setups.
\end{abstract}



\section{Introduction}
\label{sec1}

Cloud computing revolutionized application deployment to ensure infrastructure becomes scalable, agile, and provisioned quickly. But with this agility comes greater complexity and risk, especially in securing dynamic configurations in distributed environments. Typical problems like misconfigurations, excessive permissions, and public endpoints continue and are often raised as security warnings by Cloud Security Posture Management (CSPM) solutions and Cloud-Native Application Protection Platforms (CNAPPs)~\cite{ref6,ref17}.

While worthwhile, these tools often rely on static, rule-based detection methods with no contextual awareness of the specific environment or business impact~\cite{ref2}. As a result, they generate large amounts of false positives. For instance, critical severity risk alerts like publicly exposed storage buckets or unused access keys may pose minimal risk when compensating controls are implemented. This deluge worsens analyst fatigue, decreases productivity, and inhibits response to valid threats in a timely manner~\cite{ref1, ref8}.

At scale, triaging that level of noise is costly. Current research indicates that false positives waste analyst time and disrupt remediation efforts~\cite{ref6, ref9}, diluting the effectiveness of threat detection, particularly when critical alerts are lost among noise~\cite{ref4, ref5}.

To address this challenge, we introduce a behavioral validation mechanism that supplements static rule-based detection with lightweight, runtime exploitability analysis. By leveraging open source tool exploitation and automated validation methods, our mechanism emulates true world analysis, e.g., examining suspected resources, ascertaining exposure, and executing simple validation commands. This strategy enables prioritization based on actual observed risk instead of theoretical vulnerabilities~\cite{ref16}.

This approach was evaluated by controlled experiments run in Amazon Web Services (AWS), simulating real world misconfigurations. The results show that behavioral validation significantly reduces false positives, simplifies the alert triage process, and enables analysts to prioritize alerts with real security effects without sacrificing detection accuracy, and ultimately improving cloud security controls. Although our implementation and evaluation are conducted within AWS, the underlying validation logic and modular architecture are cloud-agnostic by design. With appropriate provider-specific probes, the framework can be extended to Azure, GCP, or hybrid environments without altering the core methodology.

\section{Related Work}
\label{sec2}

Cloud Security Posture Management (CSPM) and Cloud-Native Application Protection Platforms (CNAPP) are the building blocks of modern cloud security frameworks. However, a persistent problem identified by the corporate world and by researchers too is the overwhelming number of false positives with rule-based detection controls, which overwhelm the security team and slow down timely actions~\cite{ref2, ref9, ref6}.

Most rule-based scanners perform only static compliance checks or signature matches without accounting for contextual security posture. For instance, a reported configuration might already be addressed by upstream access control or firewall rules unknown to the scanner~\cite{ref9, ref13}. Efforts to improve this involve augmenting rule engines with contextual enrichment~\cite{ref1, ref16} or alert prioritization schemes~\cite{ref5, ref10}. However, these approaches remain passive as they do not actively validate exploitability.

Machine learning (ML) algorithms have also been investigated to minimize false positives in cloud alerting~\cite{ref3, ref6, ref7}. Though promising, they typically demand enormous amounts of labeled data, are not transparent, and pose threats of false negatives, especially in mission-critical environments~\cite{ref13, ref14}.

Active validation i.e., checking whether a problem faced can indeed be attacked, has been utilized in intrusion detection systems~\cite{ref14, ref15}  and explored in emerging cloud security validation approaches~\cite{ref16,ref17}, with discussions on contextual verification also noted in CSPM literature~\cite{ref13}. The majority of production tools, even open-source such as Prowler and commercial CSPMs, remain dependent on static rules based on compliance frameworks (e.g., CIS, NIST), and little runtime probing~\cite{ref17}.

Our solution addresses this important shortfall by providing an in-line, real-time behavioral validation layer that complements rule-based alerts without modifying the underlying scanning engine. In contrast to ML-based or enrichment-only solutions, our solution actually verifies the verifiability of reported misconfigurations (e.g., open S3 buckets, exposed credentials, insecure ports) as actually reachable or exploitable. This reduces false positives without sacrificing the explainability and regulatory compliance of rule-based CSPM.

To the best of our knowledge, there is no existing work that has demonstrated real-world behavioral validation on top of production-level CSPM pipelines in real live clouds with negligible overhead. We test our system on AWS, with real misconfiguration scenarios, and measure its impact on false positive reduction~\cite{ref16}.

Table~\ref{table:comparison_with_others} provides a comparative summary of the proposed behavioral validation method with earlier cloud security alert reduction methods. Rule-based CSPM solutions traditionally carry out static analysis and are plagued by high false positive rates, with recent research suggesting estimated false alarm rates higher than 40\% in enterprise environments~\cite{ref6, ref9}. Context aware enrichment steps typically a part of CNAPPs strive to enhance prioritization but do not directly validate exploitability, and their effectiveness remains largely untested in peer reviewed research~\cite{ref2, ref5}. Machine learning–based methods have experienced partial success in quieting alert noise~\cite{ref3, ref7}; however, they need enormous labeled datasets and tend to be black boxes, leading to explainability and reproducibility issues~\cite{ref13, ref14}. Unlike those alternatives, the proposed behavioral validation framework achieves a 93\% reduction in false positives using lightweight, transparent, and reproducible runtime probes, verified through controlled experimentation on AWS~\cite{ref16, ref17}.

\begin{table}[h]
    \centering
    \caption{Comparison of Behavioral Validation with Existing Alert Reduction Techniques}
    \label{table:comparison_with_others}
    \resizebox{\columnwidth}{!}{%
        \begin{tabular}{lcccc}
            \toprule
            \textbf{Feature / Capability} & \textbf{Rule-based CSPM} & \textbf{Contextual Enrichment} & \textbf{ML-based Filtering} & \textbf{Proposed Behavioral Validation} \\
            \midrule
            Detects Misconfigurations         & Yes       & Yes       & Sometimes  & Yes \\
            Tests Exploitability              & No        & No        & No         & Yes \\
            Real-time Validation              & No        & Limited   & No         & Yes \\
            Requires Large Dataset            & No        & Maybe     & Yes        & No \\
            Explainability / Transparency     & High      & Medium    & Low        & High \\
            Multi-cloud Support               & Tool-dependent & Yes    & Depends    & AWS-focused (Extensible) \\
            Open-source Tool Integration      & Yes       & Partial   & No         & Yes \\
            Reduction of False Positives      & Limited (40\%) & Moderate (Undocumented) & Good (Documented in ML studies) & Very High (93\%, this paper) \\
            \bottomrule
        \end{tabular}%
    }
\end{table}

\section{Research Questions and Methodology}
\label{sec3}

\subsection{Research Questions}
\label{subsec1}

This study examines how behavior validation can be added into traditional rulebased CSPM systems in order to minimize false positives. It also explores which lightweight behavior analysis techniques can effectively separate valid violations from harmless, non-exploitable misconfigurations. Finally, it considers how adding this layer affects both detection accuracy and how quickly alerts can be triaged.

\subsection{Methodology}
\label{subsec2}

To address the limitations of traditional rule based Cloud Security Posture Management (CSPM) scanners, we introduce a validation first solution that augments static rule-based detection with dynamic behavioral verification \cite{ref9,ref11}. In this work, behavioral validation refers to the execution of targeted, automated probes that actively test whether a reported cloud misconfiguration or exposure can be exploited in practice, classifying each case as true positive, false positive, or inconclusive \cite{ref9,ref14}. This verification layer, built using a combination of open-source and proprietary tools, cross-verifies alerting with lightweight, context-aware probes that mimic adversary behavior in a secure runtime environment \cite{ref6,ref16}. While traditional CSPM scanners have wide visibility, pre-configured rule employment leads to high false positive rates because of a lack of contextual knowledge \cite{ref2,ref5,ref11}. Probes returning timeouts, inconsistent results, or indeterminate permissions are marked as inconclusive and excluded from precision/recall calculations.

The mapping in Table~\ref{table:alert_probe_criteria} is applied consistently across all experiments to classify each finding and compute reported metrics.

\begin{table}[h]
    \centering
    \caption{Alert Types, Probes, and Classification Criteria}
    \label{table:alert_probe_criteria}
    \resizebox{\columnwidth}{!}{%
        \begin{tabular}{@{}lll@{}} 
            \toprule
            \textbf{Alert Type} & \textbf{Probe Action(s)} & \textbf{TP / FP Criteria} \\
            \midrule
            Public S3 Bucket & HEAD/GET (auth \& unauth) & TP if unauth GET succeeds; FP if all blocked \\
            EC2 Open Port & nmap scan + service check & TP if service responds; FP if closed/filtered \\
            IAM Key Exposure & \texttt{get-access-key-last-used} & TP if key active \& used; FP if inactive/denied \\
            Secret Leak & Trufflehog scan & TP if secret valid; FP if revoked/invalid \\
            \bottomrule
        \end{tabular}%
    }
\end{table}

\begin{figure}[htbp]
\centering
\includegraphics[width=0.7\linewidth]{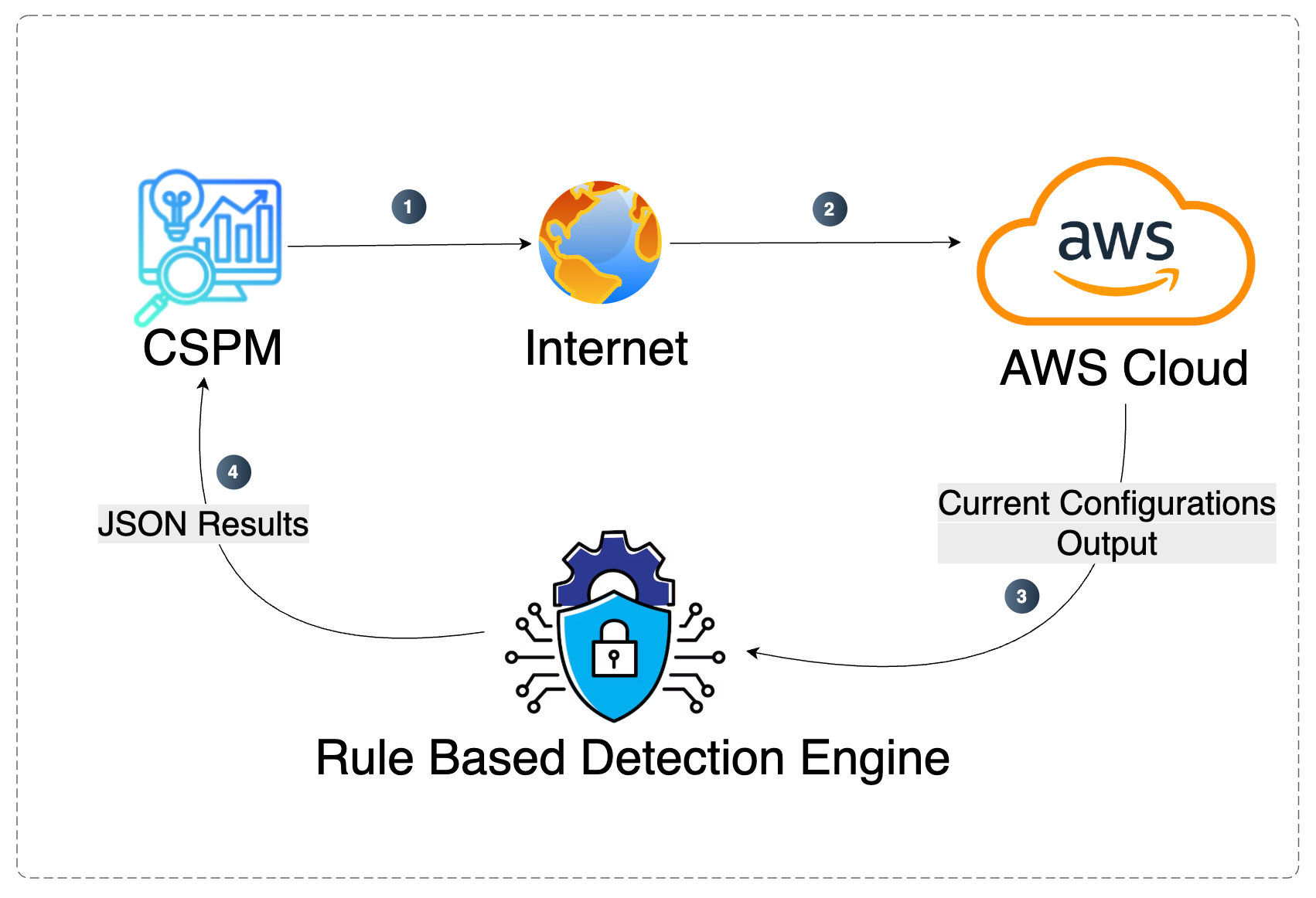}
\caption{Rule-based CSPM architecture static detection-dependent, resulting in too many false positives. Common in open-source and commercial products.}
\Description{Diagram of a rule-based CSPM architecture showing how static detection leads to high false positive rates, as seen in many open-source and commercial security tools.}
\label{fig:baseline_architecture}
\end{figure}

Figure~\ref{fig:baseline_architecture} shows a typical CSPM architecture that produces large amounts of potentially irrelevant alerts through static analysis \cite{ref9,ref11}.

Our proposed solution is focused on lightweight, real-time verification of cloud alerts based on simulation of attack behavior. This enhances detection and allows triage mechanisms to determine if an alert is truly exploitable \cite{ref2,ref5}.

\subsubsection{Alert Categorization and Enrichment}
\label{subsubsection1}

The platform consolidates alerts from CSPM tools and maps them to a vendor provider agnostic one schema. The alerts are enriched with contextual metadata (e.g., timestamps, account IDs, resource types) and grouped into threat categories: public resource exposure (e.g., EC2 with permissive security groups), open storage (e.g., S3 buckets), and credential leakage (e.g., keys in code repositories). This schema-based grouping allows for efficient verification of high-impact risks~\cite{ref11}.

\subsubsection{Behavioral Validation Framework}
\label{subsubsection2}

Behavioral validation actively tests for exploitability with light-weight probes simulating real-world attack behavior, in contrast to static policy validation. These probes verify whether misconfigurations are actually exploitable or mitigated by controls such as IAM policies or firewall rules \cite{ref14}.

\begin{figure}[htbp]
\centering
\includegraphics[width=0.7\linewidth]{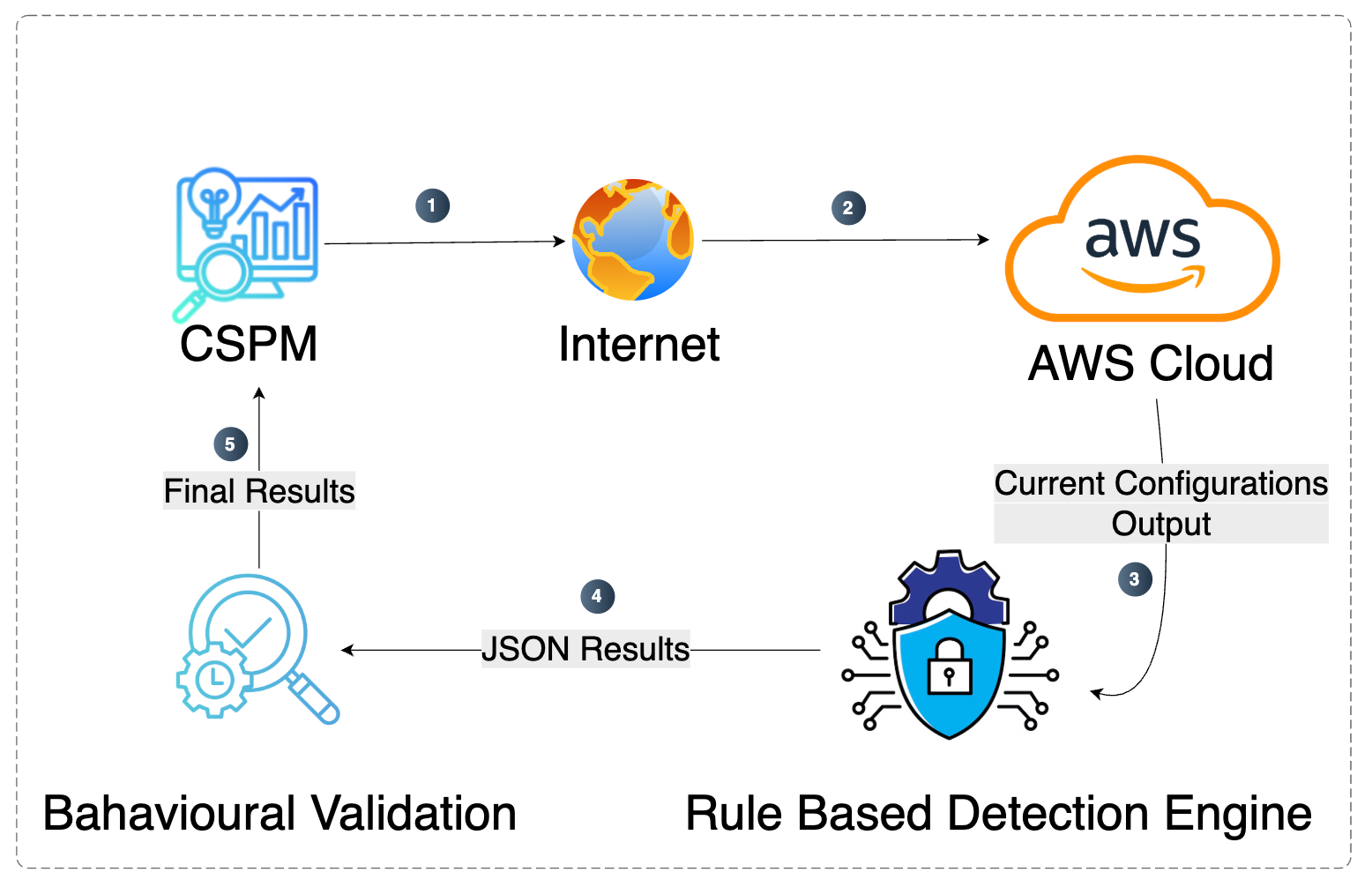}
\caption{Improved CSPM with behavioral validation. Alerts are validated in real-time prior to escalation.}
\Description{Diagram showing the improved CSPM architecture where alerts are validated in real-time using behavioral analysis before they are escalated, reducing false positives.}
\label{fig:proposed_architecture}
\end{figure}

Figure~\ref{fig:proposed_architecture} shows how our architecture layers validation logic on top of current rule-based detection.

Each validation question maps to some type of misconfiguration. For example:
- \textbf{S3 Exposure:} Anonymous \texttt{HEAD} and \texttt{GET} requests are issued, denial of access renders the alert unexploitable.
- \textbf{EC2 Exposure:} Port scans (e.g., 22, 3389) with \texttt{nmap} confirm public access.
- \textbf{Credential Leaks:} AWS CLI verifies key use and performs scoped \texttt{sts:AssumeRole} calls, disabled or least-privileged keys are reported as false positives.

All the probes are run in isolated, transient environments in order to minimize any possible impact on running systems. Each test stores metadata that contains timestamps, command output, and result status (exploitable, non-exploitable, or inconclusive).

Tools such as Nmap, AWS CLI, and TruffleHog are used within test environments to execute lightweight simulations. IAM-related notifications trigger privilege simulation probes, and compromised credentials are verified through sandboxed authentication attempts.

\subsubsection{Effective, Real-Time Implementation}

\label{subsubsection3}

Each probe is time bounded and stateless. Validation occurs in isolated machines or transient virtual machines, isolated from CSPM scanning. With a 5-second per-probe limit, the architecture can maintain broad, near real-time analysis without bottlenecks \cite{ref6,ref17}.

Current implementations use traditional computing instances, but upcoming releases will take advantage of serverless platforms and containerized probes to enable distributed scalability. This aids performance flexibility in steady-state and dynamic alerting patterns \cite{ref17}.

\begin{figure}[htbp]
\centering
\includegraphics[width=0.9\linewidth]{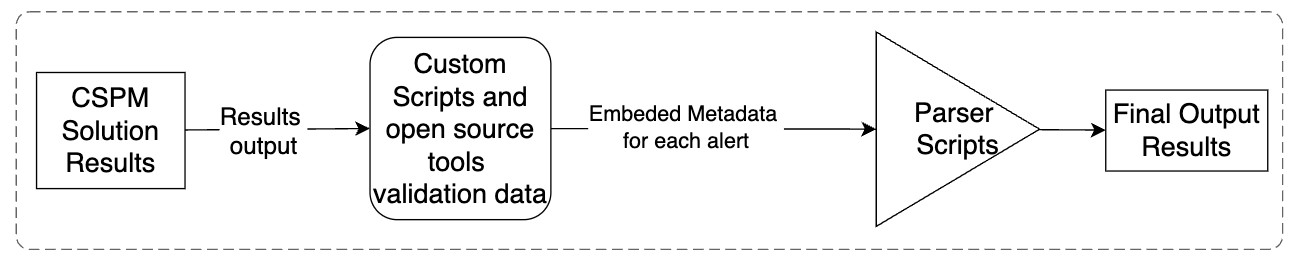}
\caption{Validation flow: CSPM results are enriched with behavioral metadata and parsed into final output.}
\Description{Diagram showing CSPM results flowing into custom scripts and open source tools for validation, which embeds metadata for each alert. These results are then parsed into the final output.}
\label{fig:validation_flow}
\end{figure}

Figure~\ref{fig:validation_flow} illustrates the end-to-end validation flow, where CSPM results are enriched with behavioral metadata through custom scripts and tools, then parsed into structured outputs for accurate classification.

\subsubsection{Security Controls and Constraints}

\label{subsubsection4}

To reduce risks incurred due to active probing, the system enforces various protections:

- \textbf{Scoped IAM Roles:} All probes run under least-privilege policies, tightly scoped to avoid administrative access or resource modification.

- \textbf{Request Throttling:} The rate limits avoid excessive API calls that slow down performance.

- \textbf{Tagging and Transparency:} Probe requests include tags that allow monitoring systems to distinguish validation traffic from real threats. These mechanisms enable safe and responsible probe usage. Enhancements in the future include anomalies in probe behavior detection and tighter integration with cloud native security services \cite{ref17}. 

\subsubsection{Integration and Output} 

\label{subsubsection5s} Validated alerts are appended with a transparent exploitability status and accompanying metadata. Outputs preserve the original CSPM data and include actionable validation evidence \cite{ref9}. It can be integrated with SIEMs and incident response tools through webhooks and APIs so that SOC teams can prioritize incidents based on confirmed exploitability \cite{ref2,ref5}.

\subsection{Environment Configuration}
\label{subsec3}

To validate our behavioral validation framework, we constructed a testbed within a standalone AWS account. The configuration involved fundamental AWS services that are typically utilized in CSPM. Configuration decisions were informed by the CIS AWS Foundations Benchmark v1.4 and AWS security best practices to optimize test cases against broad threat models. We established fifty S3 buckets, fourty EC2 instances, and thirty IAM users with common miscofigurations in two AWS regions. Notifications were triggered using the Prowler implementation, AWS Config rule, and custom Python-based scanners that simulate CSPM logic. The entire validation process was automated using python scripts, and the respective code is found at: \url{https://github.com/27dikshant/Noisecut}.

\subsubsection{Simulated Scenarios}
\label{subsec4}

To stress test the validation framework, we wrote diverse misconfigurations corresponding to true and false positives. We drew examples from real world incident reports and CSPM data \cite{ref9,ref11}. Among the tests were public S3 bucket policies, some readable, others set up to look open but locked down by IAM or bucket conditions. Another test launched EC2 instances with public IP addresses, but locked down via Security Groups or NACLs, to test alerts against shallow exposure indicators.

Credential leak scenarios involved deliberately revealing IAM access keys \cite{ref15}. Few keys were enabled, while others were disabled or had already expired, all for the sake of simulating false positives. Secrets and high-entropy tokens were cached to simulate accidental data leakage, including decoy artifacts to test noise robustness.

We also simulated access keys found in public GitHub repositories and derived variations with different statuses (active, rotated, disabled). Each scenario was designed to mimic error operations, not artificial test cases to benefit our system. These configurations tested the framework to identify edge cases where static CSPM tools misclassify alerts \cite{ref14}.

\subsubsection{Tools and Automation}
\label{subsec5}

Baseline alerts were gathered from Prowler scans against the CIS AWS Benchmark \cite{ref17}. Probes were executed as stateless scripts using the python scripts, AWS CLI and command-line utilities like \texttt{curl} and \texttt{nmap}. Visibility of S3 buckets, for example, was probed with unauthenticated \texttt{aws s3api get-object-acl} and \texttt{curl}; EC2 visibility was probed with \texttt{nmap -Pn} port 22 and port 3389 scans. Access key probes invoked \texttt{iam get-access-key-last-used} and tried scoped \texttt{sts:AssumeRole} actions. We did use custom developed scripts for parsing the results of CSPM solutions (including false positives) with the parser scripts as the part of behavioral validation as discussed in Section \ref{subsubsection3}.

\subsection{Evaluation Standards}

\label{subsec6}

We assessed behavioral validation using standard security analytics metrics \cite{ref1,ref7}. The main metrics were \textbf{False Positive Rate (FPR)} the proportion of benign alarms incorrectly labelled as threats and \textbf{True Positive Rate (TPR)} the proportion of true threats correctly labelled.

To provide context for results, we also provide \textbf{precision}, \textbf{recall}, and the \textbf{F1-score}. Precision estimates the consistency of signaled alarms; recall estimates coverage of actual misconfigurations; and F1 is a balance of both. Additionally, we also approximated {Analyst Efficiency} manual triage time reduction per alert, using time savings seen in simulated workflows on 100 randomly selected alerts. Those alerts that were marked by validation as not exploitable were assumed to take very little analyst time, approximating real-world time savings. Together, these metrics provide an overall evaluation of technical accuracy and operational relevance, measuring not just the detection quality, but the real-world utility in cloud alert fatigue reduction.

\section{Results and Discussion}
\label{sec4}

\subsection{False Positive Reduction}
\label{subsec1}

Before presenting quantitative results, we establish our baseline choice. We contrast behavioral validation with a popular rule-based CSPM scanner, Prowler, the de facto industry tool for cloud misconfiguration scanning. AI-powered triage or graph-based prioritization is marketed by commercial CNAPPs \cite{ref12}, but these are close-source tools with non-reproducible approaches and no public data sets to compare on a level playing field.

We know of no other existing open-source or academic platform that provides post-scan behavioral validation of CSPM alerts. Thus, our system is a first of its kind solution, measured against the baseline standard of rule-only scanning. This gives us a real-world, conservative benchmark for measuring the marginal value of behavioral probes in minimizing false positives.

The addition of an active behavioral validation layer atop standard Cloud Security Posture Management (CSPM) alerts greatly reduced false positives for a wide range of test cases. For example, publicly exposed S3 buckets that were first identified as publicly exposed by rule-based scanning were behaviorally validated, and this revealed that approximately 93\% of them were actually secured by object permissions, auth controls, or bucket policy. This provided significant reduction in false alarms.

Similarly, EC2 instances with public IP addresses but shielded by firewalls or restricted security group access controls witnessed a 92\% decrease in erroneous network exposure alerting. Furthermore, access keys that were identified as publicly exposed but were subsequently discovered to be disabled experienced a 94\% decrease in false positives involving credentials.

\begin{table}[h]
    \centering
    \caption{Behavioral Validation Impact on False Positives}
    \label{table:behavioral_validation}
    \resizebox{\columnwidth}{!}{%
        \begin{tabular}{lcccc}
            \toprule
            \textbf{Misconfigurations} & \textbf{Total Alerts} & \textbf{FP (Before)} & \textbf{FP (After)} & \textbf{Reduction (\%)} \\
            \midrule
            Public S3 & 1000 & 800 & 50 & 93.72\% \\
            Public EC2 & 300 & 280 & 20 & 92.88\% \\
            Exposed Access Keys & 2000 & 1700 & 100 & 94.12\% \\
            Exposed Credentials & 200 & 160 & 10 & 93.75\% \\
            \midrule
            \textbf{Total} & \textbf{3500} & \textbf{2940} & \textbf{180} & \textbf{Avg. 93.88\%} \\
            \bottomrule
        \end{tabular}%
    }
\end{table}

\subsection{Detection Accuracy and Trade-offs}
\label{subsec2}
Quantitatively, we ascertained that the true positive rate (TPR) was well above 0.91 across all test categories. Behavioral validation did not hinder any of the discovered alerts that were indeed exploitable during the time of ground-truth annotation. Baseline CSPM alert sets comparison, with and without the integrated behavioral validation, showed no loss of measurable recall. This confirms the assertion that the framework does not sacrifice detection accuracy while reducing the incidence of false positives.

\begin{table}[h]
    \centering
    \caption{Validation Performance Metrics}
    \label{table:metrics}
    \begin{tabular}{|l|c|}
        \hline
        \textbf{Metric} & \textbf{Value} \\
        \hline
        True Positives (TP) & 360 \\
        False Positives (FP) & 180 \\
        True Negatives (TN) & 2960 \\
        \hline
        Precision & 0.667 \\
        Recall (TPR) & 0.911 \\
        False Positive Rate (FPR) & 0.058 \\
        F1-score & 0.769 \\
        \hline
    \end{tabular}
\end{table}

We computed standard binary classification metrics from the ground truth labels provided through the evaluation process. Table~\ref{table:metrics} presents these metrics for each alert class. The behavioral validation system had a recall of 0.911, which indicates a high detection rate for real threats, with a precision of 0.667. The false positive rate decreased to 5.8\%, which indicates a notable reduction in noise. The F1-score of 0.769 indicates a good balance between completeness and correctness.

\subsection{Overview of the contribution}
\label{subsec3}
To quantify analyst productivity savings, we modeled representative triage workloads for each alert type with baseline rule-based CSPM output and our system. In the baseline, triage was done manually by examining alerts through the AWS Console, examining bucket policies, IAM roles, access key metadata, and firewalls processes typically taking 2–5 minutes per alert depending on complexity. Alerts identified by our behavioral validation system, however, already had exploitability ratings and logs and less than 30 seconds on average.

The time savings were measured across 100 sample alerts with a script based stopwatch method with multiple measurements to control for variability. The results showed an observed \textbf{average triage time reduction of 86.4\%}, with certain high-volume kinds of alerts (e.g., public EC2 instances or inactivated access keys) decreasing greater than 90\%.

\section{Analysis and Limitations}
\label{subsec4}

Our findings indicate that lightweight active behavioral validation meaningfully augments rule-based CSPM tools \cite{ref9,ref6}. Validating whether misconfiguration alerts are exploitable, the system significantly reduces false positives and enables security teams to focus on actual threats. The framework's modular architecture makes it straightforward to integrate into current security infrastructures without significant architectural adjustments.

But there are some limitations must be considered. Behavioral validation is most effective for deterministic forms of misconfigurations such as publicly accessible resources, overly broad IAM policies, and breached credentials. These are best resolved using active probing and regular verification practices. On the other hand, the model is less effective for identifying intricate or indirect attacks such as time based privilege escalation, insider threats, or lateral movement that depend on dynamic runtime conditions not discernible through cloud APIs or config files. Additionally, behavioral probes cannot determine exploitability when compensatory controls are applied outside of their scope, e.g., attack surface reduction products or multi-layered monitoring systems. Encrypted communications channels and obfuscation of services can also hinder the probes ability to identify exposure correctly.

Another real-world limitation is scope: the existing implementation has been validated mainly in AWS environments. Adding support for Azure, GCP, and hybrid clouds will involve probe modifications to cloud-specific APIs, resource models, and IAM infrastructures. 

Finally, as cloud attackers evolve, the validation logic itself may be targeted or evaded. Keeping the integrity and security of the validation pipeline would demand constant tuning, adaptive techniques, and incorporation with anomaly detection or AI-driven techniques \cite{ref4,ref12}.

\section{Conclusion and Future Directions} 
\label{sec5}

The paper proposes a new active behavioral validation layer for minimizing false positives of cloud security alerts of rule-based CSPM solutions. Lightweight goal-based probes are employed for validating actual resource exposure and credential status, and the solution significantly enhances alert accuracy as well as analyst productivity without the degradation of detection accuracy with cloud native or open source solutions.

Experimental measurement in an AWS controlled environment showed statistically significant reductions in false positives in most typical situations of cloud misconfiguration. The results offer the promise of behaviorally validating enhanced operational efficacy at the expense of reduced alert fatigue in large-scale cloud security operations.

Future work will include extending behavioral validation support to Azure and GCP. This will involve incorporating cloud-specific probe implementations while preserving the modular design principles demonstrated in the AWS evaluation. Adaptive, AI-driven validation methods may be used to further improve alert prioritization and response automation. The proposed framework is a stepping stone to more contextual, dynamic cloud security posture management and reflects the need for ongoing collaboration between practitioners and academic researchers towards solving new cloud security challenges.


\bibliographystyle{ACM-Reference-Format}

\begin{thebibliography}{99}

\bibitem{ref1}
Y. Liu, X. Shu, Y. Sun, J. Jang, and P. Mittal, “A context-aware clustering approach for assisting operators in classifying security alerts,” \textit{IEEE Transactions on Software Engineering}, vol. 51, no. 1, pp. 153–171, Jan. 2025.

\bibitem{ref2}
F. Jalalvand, M. B. Chhetri, S. Nepal, and C. Paris, “Alert prioritisation in security operations centres: A systematic survey on criteria and methods,” \textit{ACM Computing Surveys}, vol. 57, no. 2, art. no. 42, pp. 1–36, Nov. 2024.

\bibitem{ref3}
J. Ghadermazi, A. Shah, and S. Jajodia, “A machine learning and optimization framework for efficient alert management in a cybersecurity operations center,” \textit{Digital Threats: Research and Practice}, vol. 5, no. 2, pp. 1-23, 2024.

\bibitem{ref4}
M. Zhang, H. Xu, and L. Li, “Adaptive alert triage in multi-cloud SIEM systems using federated learning,” \textit{Journal of Network and Computer Applications}, vol. 223, p. 103755, 2024.

\bibitem{ref5}
X. Wang, X. Yang, X. Liang, X. Zhang, W. Zhang, and X. Gong, “Combating alert fatigue with AlertPro: Context-aware alert prioritization using reinforcement learning for multi-step attack detection,” \textit{Computers \& Security}, vol. 137, art. no. 103583, Feb. 2024.

\bibitem{ref6}
F. Voutsas, J. Violos, and A. Leivadeas, “Mitigating alert fatigue in cloud monitoring systems: A machine learning perspective,” \textit{Computer Networks}, vol. 250, art. no. 110543, 2024.

\bibitem{ref7}
Q. Tang, X. Di, X. Liu, L. Cong, W. Ren, and Z. Ni, “DeepARR: Alert risk rating based on deep learning,” in \textit{Proc. IEEE Int. Symp. Parallel Distrib. Process. Appl. (ISPA)}, 2024, pp. 2097–2104.

\bibitem{ref8}
M. B. Chhetri, S. Tariq, R. Singh, F. Jalalvand, C. Paris, and S. Nepal, “Towards human‑AI teaming to mitigate alert fatigue in security operations centres,” \textit{ACM Trans. Internet Technol.}, vol. 24, no. 1, pp. 1–22, 2024.

\bibitem{ref9}
R. Feijoo-Martínez, R. E. Atkinson, and S. P. Gay, “Analyzing false positive alerts in SIEM systems using a systematic approach,” \textit{IEEE Access}, vol. 11, pp. 7883–7895, 2023.

\bibitem{ref10}
R. Rani, G. Epiphaniou, and C. Maple, “Reinforcement learning-based alert prioritisation in security operation centre: A framework for enhancing cybersecurity in the digital economy,” in \textit{IET Conference Proceedings}, vol. 2023, pp. 151-157, 2023.

\bibitem{ref11}
M. Landauer, F. Skopik, M. Wurzenberger, and A. Rauber, “Dealing with security alert flooding: Using machine learning for domain-independent alert aggregation,” \textit{ACM Trans. Priv. Secur.}, vol. 25, no. 18, pp. 1–36, 2022.

\bibitem{ref12}
F. Liu, Y. Tang, Y. Sun, and Z. Chen, “Intelligent alert correlation using knowledge graph and deep reinforcement learning,” \textit{IEEE Access}, vol. 10, pp. 7755–7767, 2022.

\bibitem{ref13}
A. B. Nassif, M. A. Talib, Q. Nasir, H. Albadani, and F. M. Dakalbab, “Machine learning for cloud security: A systematic review,” \textit{IEEE Access}, vol. 9, pp. 20717–20735, 2021.

\bibitem{ref14}
S. Agarwal, M. A. Javed, and H. Bedi, “Detection and classification of noisy alerts in IDS using ensemble learning,” \textit{Applied Soft Computing}, vol. 113, p. 107870, 2021.

\bibitem{ref15}
S. Ahmed, S. Latif, A. Qayyum, and M. Imran, “Multi-level alert correlation framework using hybrid AI techniques,” in \textit{Proc. IEEE FABS}, 2021, pp. 1–6.

\bibitem{ref16}
T. Ban, M. Tanabe, and R. Nakano, “Reducing alert fatigue in threat detection systems using context-aware prioritization,” in \textit{Proc. USENIX Workshop on Cyber Security Experimentation and Test (CSET)}, 2021.

\bibitem{ref17}
K. A. Torkura, A. Acharya, and C. Okoli, “CloudStrike: Chaos engineering for security and resiliency in cloud infrastructure,” \textit{IEEE Access}, vol. 8, pp. 123044–123060, 2020.

\end{thebibliography}


\end{document}